# Equivalence of two alternative approaches to Schrödinger equations

B. Gönül and K. Köksal

Department of Engineering Physics, University of Gaziantep, 27310, Gaziantep-Türkiye

**Abstract**
Recently developed simple approach for the exact/approximate solution of Schrödinger equations with constant/position-dependent mass, in which the potential is considered as in the perturbation theory, is shown to be equivalent to the one leading to the construction of exactly solvable potentials via the solution of second-order differential equations in terms of known special functions. The formalism in the former solves difficulties encountered in the latter in revealing the corrections explicitly to the unperturbed piece of the solutions whereas the other obviate cumbersome procedures used in the calculations of the former.

Using the spirit of prescriptions used for the usual treatments in supersymmetric quantum theory [1], a simple alternative approach has been recently developed [2] to perturbation theory in one-dimensional non-relativistic quantum mechanics. The formulae proposed in this work for the energy shifts and wave functions do not involve tedious calculations, unlike the standard perturbation theory, where it has been clearly shown that the supersymmetric perturbation techniques [3] and other approaches [4] based logarithmic perturbation theory [5] are involved within the more general framework of this novel formalism [2].

Further, this procedure has been used for the exact treatments of quantum states with non-zero angular momentum in case effective potentials have an analytical solution [6], for investigation of perturbed Coulomb interactions [7] and for a systematic search of exactly solvable non-central potentials [8]. The model is later extended to the scattering domain [9] deriving explicitly the changes in the partial wave phase shifts. Successful applications of the model to anharmonic oscillators and Yukawa type potentials are presented in [10] and [11], respectively . With the confidence gained from such applications involving constant mass, the model underlined is employed for the investigation of effective Hamiltonians with spatially varying mass in one-dimension [12] and higher dimensions [13] leading to explicit expressions for the energy eigenvalues and eigenfunctions. These investigations have clarified the relation between exact solvability of the effective mass Hamiltonians in the literature and ordering ambiguity put forward due to the use of a position-dependent mass in equations. The model restricts naturally the possible choices of ordering and provide us an explicit comparison betweeen physically acceptable Hamiltonians [12,13].

Within this context, the purpose of this Letter is to show that the model of interest is entirely equivalent, as has been proven first for the perturbation theories mentioned above, to the other powerful technique(s) suggested for exactly solvable Schrödinger equations with constant [14] and location dependent masses [15]. Hence, the present brief study completes the unification of the scheme in [2] with the other alternative treatments used for perturbative problems and analytically solvable potentials. Despite this equivalence, it seems that their developments in the literature have been completely independent although each turns out to resolve difficulties encountered in the other.

To understand how this happens, in case of now exact solvability, firstly we introduce a unified formalism in the following section together with two illustrations having constant and non-constant masses leading to a comparison between the results of the models interested. Further discussion on the unified treatment of both methods to illuminate and perfects the calculations is presented.

Many of the special functions $F(g)$ of mathematics represent solutions to differential equations of the form

$$\frac{d^2 F(g)}{dg^2} + Q(g)\frac{dF(g)}{dg} + R(g)F(g) = 0 , \qquad (1)$$

where the functions $Q(g)$ and $R(g)$ are well defined for any particular function [16]. Since in this Letter we are interested in bound state wave functions, we should restrict ourselves to polynomial solutions of Eq. (1). The substitution of $\Psi(x) = f(x)F[g(x)]$ in the most favourable one-dimensional time-independent Schrödinger equation [12,13,17] associated with a particle endowed with a position-dependent effective mass

$$-\frac{d}{dx}\left[\frac{1}{M(x)}\frac{d\Psi(x)}{dx}\right] + V(x)\Psi(x) = E\Psi(x) , \qquad (2)$$

leads to the second-order differential equation

$$\frac{1}{M}\left(\frac{f''}{f} + \frac{F''g'^2}{F} + \frac{g''F'}{F} + 2\frac{F'g'f'}{Ff}\right) - \frac{M'}{M^2}\left(\frac{f'}{f} + \frac{F'g'}{F}\right) = V - E , \qquad (3)$$

in which primes denote derivatives with respect to the variables and $M(x)$ is the dimensionless form of the mass function $m(x) = m_0 M(x)$ where we have set $\hbar = 2m_0 = 1$. From our earlier applications [6-13], we split Eq. (3) in two pieces

$$W^2(x) - \left[\frac{W(x)}{\sqrt{M}}\right]' = V_0(x) - \varepsilon \quad , \quad W = -\frac{f'}{\sqrt{M} f} \quad , \tag{4}$$

and

$$\Delta W^2(x) - \left[\frac{\Delta W(x)}{\sqrt{M}}\right]' + 2W(x)\Delta W(x) = \Delta V(x) - \Delta E \quad , \quad \Delta W = -\frac{F'g'}{\sqrt{M} F} \quad , \tag{5}$$

where $E = \varepsilon + \Delta E$ and $V = V_0 + \Delta V$. Eqs. (4) and (5) are the concrete proof of the equivalence between the two alternative models appeared in [2, 6-13] and [14,15] for constant and non-constant mass cases, because for $M \to 1$ the above equations reduce to the well known equations with constant mass.

Afterall, it can be clearly seen that Eq. (4) is the one required to obtain an explicit expression for $W$ term corresponding to exactly solvable systems in one-dimension. However, to proceed further, the functions $f$ and $g$ should be solved first as $F$, $Q$ and $R$ are known in principle. Now, equating like terms between the resulting expression in (3) and (1) gives

$$Q[g(x)] = \frac{1}{g'}\left(\frac{g''}{g'} + \frac{2f'}{f} - \frac{M'}{M}\right) \quad , \quad R[g(x)] = \frac{1}{g'^2}\left[\frac{f''}{f} - \frac{M'}{M}\frac{f'}{f} + M(E - V)\right] \quad , \tag{6}$$

where, from the definition of $Q$,

$$f(x) \approx \left(\frac{M}{g'}\right)^{1/2} \exp\left[\frac{1}{2}\int^{g(x)} Q(g)dg\right] . \tag{7}$$

Consideration of Eqs. (3) through (6) suggests a novel prescription

$$\Delta V(x) - \Delta E = -\frac{g'^2}{M} R(g) \quad , \tag{8}$$

which alone, for plausible $M$ and $R$ functions, provides a reliable expression for $g(x)$. Eq. (8) removes the ambiguity encountered in [14,15] while setting a correct energy term (corresponding here to $\Delta E$) in between possible choices arised naturally.

To clarify this point, we choose one simple example considered recently [15] where the mass function is $M(x) = \lambda g'(x)$ being with $\lambda$ is a positive constant. Assuming that

$F_n(g) \propto P_n^{(\alpha,\beta)}(g)$, $n = 0,1,2,...$, $\alpha, \beta \succ -1$ is either a Jacobi or generalized Laguerre polynomials [16], hence

$$R(g) = \frac{n(n+\alpha+\beta+1)}{1-g^2}, \quad Q(g) = \frac{(\beta-\alpha)}{1-g^2} - (\alpha+\beta+2)\frac{g}{1-g^2}. \tag{9}$$

A constant term corresponding to $\Delta E$ therefore appears on the right-hand side of Eq. (8) with the assumption that $g'/[\lambda(1-g^2)] = C$ where $C$ must be restricted to positive values in order to get $E = \varepsilon + \Delta E \prec 0$ for the bound states. The solution of this first-order differential equation for $g(x)$ leading to a positive mass function reads

$$g(x) = \tanh qx, \quad M(x) = \operatorname{sech}^2 qx, \tag{10}$$

where $qC \succ 0$ and $C = q^2$. Eq. (7) then yields

$$f(x) = \sqrt{\frac{1}{q}} (\tanh qx - 1)^{(\alpha+1)/2} (\tanh qx + 1)^{(\beta+1)/2}, \tag{11}$$

which, through (Eq.4), leads to

$$V_0(x) - E_0 = \frac{q^2}{4}\left[(\alpha^2-1)e^{2qx} + (\beta^2-1)e^{-2qx}\right] - \frac{q^2}{2}(\alpha+1)(\beta+1). \tag{12}$$

Therefore, an exactly solvable full potential term forms in $V(x) = V_0(x)$ as there is no contribution due to $\Delta V(x)$ in Eq. (8) for the specifically chosen example here. Then, the corresponding energy expression and wave function are

$$E = E_0 + \Delta E = \frac{q^2}{2}(\alpha+1)(\beta+1) + q^2 n(n+\alpha+\beta+1), \tag{13}$$

$$\Psi_n(x) \propto f(x) F[g(x)] = (\tanh qx - 1)^{(\alpha+1)/2} (\tanh qx + 1)^{(\beta+1)/2} P_n^{(\alpha,\beta)}(\tanh qx), \tag{14}$$

where it is obvious that $\alpha, \beta \succ -1$ in order to provide $E \prec 0$ and to satisfy conditions on bound-state wave functions. These results are in agreement with the work in [15].

For the another application of the model in case of a constant particle mass $(M \to 1)$, we focus on the attractive radial potential studied in the work of Williams and Poulious [14] where

$$R[g(x)] = \frac{(n+\alpha)^2}{1-g^2} + \frac{2+4\alpha(1-\alpha)+g^2}{4(1-g^2)^2}, \quad Q[g(x)] = 0, \tag{15}$$

in Eq. (1) assuming that $F_n(g) \propto (1-g^2)^{(\alpha+1/2)/2} G_n^\alpha[g(r)]$ related to the Gegenbauer polynomials $G_n^\alpha(g)$ [16]. Since we have to get a constant $(\Delta E)$ on the left-hand side of Eq.

(8), there must be at least one term on the right-hand side, from which a constant arises. In the most general case this must be one of the terms containing the parameters $n$ and $\alpha$ of the Gegenbauer polynomials. For this, eq. (15) is rearranged as

$$R(g) = \left\{ \frac{(n+\alpha)^2 + \alpha(1-\alpha) + (1/2)}{(1-g^2)^2} + \frac{[(1/4)-(n+\alpha)^2]g^2}{(1-g^2)^2} \right\}, \tag{16}$$

which, with the consideration of Eq. (8), puts forwards two possibilities

$$g'^2/(1-g^2)^2 = C \quad \text{and} \quad (g'^2)g^2/(1-g^2)^2 = C, \tag{17}$$

where $C = a^2$ is a positive constant in order to reproduce physically acceptable solutions. We can get different kinds of $g(x)$ functions from these differential equations. For a detail investigation on the related topic the reader is referred to the excellent works of Levai [14]. To compare our result, we choose the second one as in the work of Williams and his co-worker [14] where the solution $g(x) = [1-\exp(2ax)]^{1/2}$ which, through Eq. (8), yields

$$\Delta V(x) = -\frac{a^2 A}{[1-\exp(2ax)]}, \quad \Delta E = a^2 \left[\frac{1}{4} - (n+\alpha)^2\right], \tag{18}$$

in which $A = (n+\alpha)^2 + \alpha(1-\alpha) + (1/2)$. From Eqs. (7) and (4),

$$V_0(x) = \frac{a^2[3-6\exp(2ax)]}{4[\exp(2ax)-1]^2}, \quad E_0 = -\frac{a^2}{4}, \tag{19}$$

and ultimately

$$V(x) = V_0 + \Delta V = \frac{a^2}{4}\left[\frac{(6-4A)g^2 - 3}{g^4}\right] = \frac{a^2}{4}\left\{\frac{(3-4A)\exp(-4ax) + (4A-6)\exp(-2ax)}{[1-\exp(-2ax)]^2}\right\},$$

$$E = E_0 + \Delta E = -a^2(n+\alpha)^2, \tag{20}$$

and the corresponding wave function easily can be calculated through $\Psi(x) \propto f(x)F(g)$ if required. These results are in agreement with the related work [14] and share similarities with the $(n+1)$−member Hamiltonian hierarchy problem [18] having a super-family potential that is reduced to the Hulthen potential for the first member $(n=0)$.

The work presented in this paper through the two examples discussed above would appear to be of interest on two grounds. First, the successful unification of both model and their equivalence to each other are nicely presented, which have not been noticed before. And, the association is not an idle curiosity, for each method has much to tell the other: the model in [2,6-13] offers the explicit formulas for the energy and wave function corrections, which are absent in the other [14,15], while [14,15] provide a clean route to a compact calculation, which may be seen cumbersome for analyzing such problems in [2,6-13]. Although the procedure used in [2,6-13] may not actually reduce the calculational workload, it does cast the calculations into a physically-motivated, visualizable framework.

Second, of greater theoretical interest is the clarification that the structure of $R(g)$ via Eq. (8) helps us to understand which special functions lead to exactly solvable potentials whereas why some potentials have no analytical solutions. This point confirms the significance of Eq. (8) which reveals the solutions for a given piece of the potential term yielding directly $g(x)$, $f(x)$ and $W(x)$ terms. Along this line, considering the structures of possible $R(g)$ functions in the literature [16] and the search in [10,11] carried out for approximately solvable potentials used fequently in physics, together with the expansion of (Eq.8) employed for perturbative treatments in these references, the works are in progress. We hope that this investigation would provide us an alternative scheme in terms of special functions of mathematical physics to initiate calculations from the modification term in the potential requiring no other information, unlike the familiar perturbation theories in the literature.